\documentclass[useAMS,usenatbib,usegraphicx]{mn2e}
\usepackage{latexsym}
\usepackage{epsfig}
\usepackage{amsmath}  
\usepackage{color}
\usepackage{amsfonts}
\usepackage{pgf}
\bibpunct{(}{)}{;}{a}{,}{,}
\usepackage[colorlinks=true,linkcolor=blue,citecolor=blue, breaklinks=true]{hyperref}

\def\lm{{\ell m}}
\def\half{\frac{1}{2}}

\def\nobs{N_{\rm obs}}

\def\rfg{N_{\rm FG}}

\makeatletter
\def\ref@jnl#1{{\rmfamily#1}}%
\newcommand\aj{\ref@jnl{AJ}}%
\newcommand\araa{\ref@jnl{ARA\&A}}%
\newcommand\apj{\ref@jnl{ApJ}}%
\newcommand\apjl{\ref@jnl{ApJ}}%
\newcommand\apjs{\ref@jnl{ApJS}}%
\newcommand\ao{\ref@jnl{Appl.~Opt.}}%
\newcommand\apss{\ref@jnl{Ap\&SS}}%
\newcommand\aap{\ref@jnl{A\&A}}%
\newcommand\aapr{\ref@jnl{A\&A~Rev.}}%
\newcommand\aaps{\ref@jnl{A\&AS}}%
\newcommand\azh{\ref@jnl{AZh}}%
\newcommand\baas{\ref@jnl{BAAS}}%
\newcommand\jrasc{\ref@jnl{JRASC}}%
\newcommand\memras{\ref@jnl{MmRAS}}%
\newcommand\mnras{\ref@jnl{MNRAS}}%
\newcommand\pra{\ref@jnl{Phys.~Rev.~A}}%
\newcommand\prb{\ref@jnl{Phys.~Rev.~B}}%
\newcommand\prc{\ref@jnl{Phys.~Rev.~C}}%
\newcommand\prd{\ref@jnl{Phys.~Rev.~D}}%
\newcommand\pre{\ref@jnl{Phys.~Rev.~E}}%
\newcommand\prl{\ref@jnl{Phys.~Rev.~Lett.}}%
\newcommand\pasp{\ref@jnl{PASP}}%
\newcommand\pasj{\ref@jnl{PASJ}}%
\newcommand\qjras{\ref@jnl{QJRAS}}%
\newcommand\skytel{\ref@jnl{S\&T}}%
\newcommand\solphys{\ref@jnl{Sol.~Phys.}}%
\newcommand\sovast{\ref@jnl{Soviet~Ast.}}%
\newcommand\ssr{\ref@jnl{Space~Sci.~Rev.}}%
\newcommand\zap{\ref@jnl{ZAp}}%
\newcommand\nat{\ref@jnl{Nature}}%
\newcommand\iaucirc{\ref@jnl{IAU~Circ.}}%
\newcommand\aplett{\ref@jnl{Astrophys.~Lett.}}%
\newcommand\apspr{\ref@jnl{Astrophys.~Space~Phys.~Res.}}%
\newcommand\bain{\ref@jnl{Bull.~Astron.~Inst.~Netherlands}}%
\newcommand\fcp{\ref@jnl{Fund.~Cosmic~Phys.}}%
\newcommand\gca{\ref@jnl{Geochim.~Cosmochim.~Acta}}%
\newcommand\grl{\ref@jnl{Geophys.~Res.~Lett.}}%
\newcommand\jcp{\ref@jnl{J.~Chem.~Phys.}}%
\newcommand\jgr{\ref@jnl{J.~Geophys.~Res.}}%
\newcommand\jqsrt{\ref@jnl{J.~Quant.~Spec.~Radiat.~Transf.}}%
\newcommand\memsai{\ref@jnl{Mem.~Soc.~Astron.~Italiana}}%
\newcommand\nphysa{\ref@jnl{Nucl.~Phys.~A}}%
\newcommand\physrep{\ref@jnl{Phys.~Rep.}}%
\newcommand\physscr{\ref@jnl{Phys.~Scr}}%
\newcommand\planss{\ref@jnl{Planet.~Space~Sci.}}%
\newcommand\procspie{\ref@jnl{Proc.~SPIE}}%

\title[Foreground Maps in WMAP frequency bands]{Foreground Maps in WMAP frequency bands}
\author[T. Ghosh et. al]
{Tuhin Ghosh$^{1,2}$\thanks{E-mail: tuhin@iucaa.ernet.in},
Jacques Delabrouille$^{2}$\thanks{E-mail: delabrouille@apc.univ-paris7.fr},
Mathieu Remazeilles$^{2}$\thanks{E-mail: remazeil@apc.univ-paris7.fr},
Jean-Fran\c{c}ois 
\newauthor
Cardoso $^{2}$\thanks{E-mail: cardoso@tsi.enst.fr} and Tarun Souradeep$^{1}$\thanks{E-mail: tarun@iucaa.ernet.in} \\
$^{1}$IUCAA, Post Bag 4, Ganeshkhind, Pune-411007, India\\
$^{2}$ APC 10, rue Alice Domon et Leonie Duquet, 75205 Paris Cedex 13, France}

\begin{document}

\date{Accepted Received}

\maketitle

\begin{abstract}
This paper provides full sky maps of foreground emission in all WMAP channels, with very low residual contamination from the Cosmic Microwave Background (CMB) anisotropies and controlled level of instrumental noise. Foreground maps are obtained by subtraction of a properly filtered CMB map, obtained from linear combinations of needlet-based representations of all WMAP observations and of a 100--micron map. The error in the reconstructed foreground maps on large scales is significantly lower than the original error due to CMB contamination, while remaining of the order of the original WMAP noise on small scales. The level of the noise is estimated, which permits to implement local filters for maximising the local signal to noise ratio. An example of such filtering, which reduces the small scale noise using latitude dependent filters is implemented. This enhances significantly the contrast of galactic emission, in particular on intermediate angular scales and at intermediate galactic latitude. The clean WMAP foreground maps can be used to study the galactic interstellar medium, in particular for the highest frequency channels for which the proper subtraction of CMB contamination is mandatory. The foregrounds maps can be downloaded from a dedicated web site.

\end{abstract}


\section{Introduction}

The WMAP space mission, launched by NASA in 2001, has been primarily designed to measure the anisotropies of the Cosmic Microwave Background (CMB) emitted when the universe became transparent, at an age of about 380,000 yr. The interpretation of these observations in the context of the standard hot Big-Bang model has constrained the main parameters of the model with great accuracy \citep{2009ApJS..180..330K}.

In addition to the primordial CMB, WMAP detectors are sensitive to foreground astrophysical emission: diffuse emission from the galactic interstellar medium (ISM), emission from compact extragalactic sources (such as radiogalaxies and AGN and, to lesser extent, thermal Sunyaev-Zel'dovich (SZ) emission from clusters of galaxies). Hence, maps observed by WMAP contain each a mixture of emissions from different astrophysical processes.

The analysis of the observed maps for deriving a CMB power spectrum,
in temperature or polarization, involves some kind of foreground
cleaning, followed by masking the regions of the sky most contaminated
by foregrounds, and estimation of the angular spectrum $C_\ell^{\rm
  CMB}$ of the CMB on the masked sky.  There exist several ways to
deal with the incomplete sky coverage for power spectrum estimation
\citep{2002ApJ...567....2H,2004MNRAS.349..603E,PhysRevD.78.083013}.
Little is lost in terms of power spectrum estimation accuracy by masking a small
fraction of sky (the accuracy is typically reduced by a factor $f_{\rm
  sky}$ equal to the fraction of the sky kept for the analysis).
Proper estimation of the contamination (biasing) of the power spectrum
by noise or residual foreground emission permits to correct for the
contribution of both kinds of contaminants on the power spectrum of
the CMB map.

Where to give up foreground cleaning and resort, instead, to masking,
is an interesting (and debated) issue.  To be on the safe side,
conservative masking is typically used for estimating the CMB power
spectrum.  However, low--level foreground emission often remains in the masked
sky (see, e.g., \cite{2005MNRAS.364.1185P,2008A&A...491..597L}).

Foreground emissions of various astrophysical origins, however, are of
much scientific interest themselves.  Scientific investigation of
foregrounds cannot resort to masking regions contaminated by CMB, the latter
being present everywhere on the sky.  In recent observations, such as
those of WMAP, the CMB is stronger than the noise on a large range of
angular scales, in all regions of the sky.  The analysis of foreground
emission in WMAP data, hence, requires subtraction of the CMB
contaminant (which is the main source of error) while keeping
contamination by instrumental noise as small as possible. Obtaining
such clean foreground intensity maps from WMAP data is the objective
of the present paper.

\section{Method}

Standard methods for component separation (see \cite{2007astro.ph..2198D} for a review) often assume that the data are well represented by noisy linear mixtures of well-defined components, e.g. CMB, SZ, thermal dust, synchrotron, etc. Most methods used in the CMB context such as Wiener filtering \citep{1999NewA....4..443B,1996MNRAS.281.1297T}, Maximum Entropy Methods \citep{1998MNRAS.300....1H}, Independent Component Analysis (ICA) methods of different types \citep{2002MNRAS.334...53M,2003MNRAS.346.1089D} explicitly represent sky emission as a superposition of emissions, the properties of which are parametrised in some way. In reality, matter emits via a great variety of distinct processes, and there is no obvious way of separating the total emission into distinct components. It is probably fair to argue that there is, for instance, no natural preferred option between distinguishing components via their emission process (e.g. synchrotron vs. free-free) or via their place of origin (the ISM of our galaxy vs. the ISM of other galaxies). Similarly, dust emission arises from a variety of forms of matter (from molecules to large dust grains), and from several emission processes (thermal greybody emission, electric dipole emission from rotating dust grains...). Whether to distinguish thermal dust from spinning dust as two separate components, or instead distinguish between warm and cold dust, for instance, is a matter of taste. One can also consider dust emission as being all emission from all matter that is not ionised -- the appropriate modeling depends on the scientific question addressed. The obvious exception to this is the CMB, which is distinct both by its origin (the last scattering surface) and its spectral emission law. If, however, the WMAP experiment was sensitive enough for kinetic SZ effect (which has the same spectral emission law as primordial CMB fluctuations) to be above the noise level, the question of whether to distinguish the two would become relevant as well.

\subsection{The model}

Given that complexity of foreground emission, and the relative lack of spectral resolution in WMAP data (only 5 channels, for many different emission processes), one must extend the way to parametrise foreground emission. Extensions of the Spectral Matching ICA (SMICA) method \citep{2003MNRAS.346.1089D} to allow more flexible models of foreground emission are discussed by \citet{2008ISTSP...2..735C,2008arXiv0803.1814C}. 

In the present analysis, we adopt the extreme point of view in which we choose to distinguish between three contributions to the WMAP maps only: CMB, foregrounds (dominated by the emission of the galactic ISM), and noise. The task we address, then, is to produce the best possible maps of foregrounds with minimal contamination from CMB and noise. Further analysis of the foreground component into different emissions of interest (whether by origin, or by emission process) is not the aim of this paper.

The temperature map at a given frequency channel of WMAP is given by
\begin{equation}
  x^i(p)= \int b^i(p.p') \left [ s(p')+ f^i(p') \right ] \,dp' + n^i(p),
\label{eq:obs-map}
\end{equation}
where $i$ indexes the WMAP frequency bands, and runs from 1 to 5. $s(p)$ and $f^i(p)$ are respectively the CMB and foreground components, and $n^i(p)$ is the instrumental noise in the given frequency band. Here, $b^i(p.p')$ is the beam of the observed map for the $i^{th}$ frequency band (assumed to be symmetric, but not necessarily Gaussian). The direction on the sky (or pixel of a pixelized map) is indexed by $p$.
Equation \ref{eq:obs-map} can be written in harmonic space as:
\begin{equation}
  x^i_\lm=  b_l^i \left [ s_\lm+ f^i_\lm \right ]  + n^i_\lm
\label{eq:obs-alm}
\end{equation}
Our objective is to estimate in the best possible way the foreground emission $\int b^i(p.p') f^i(p')dp' $ (or equivalently $b_l^i f^i_\lm$) in each WMAP channel.
Once stated in that way, the problem becomes quite well posed. The CMB emission has the particular feature that it is completely uncorrelated from the other two, and that its emission law is known (it is the derivative of a blackbody with respect to temperature, at temperature $T_{\rm CMB} = 2.725\,$K). The instrumental noise is well characterised. It is well approximated as uncorrelated from channel to channel. Its level is  known with excellent accuracy, and it is known to dominate at high angular frequencies (high $\ell$). 

The foreground component is everything else, i.e. everything that comes from the sky (rather than from the instrument) and is not CMB. We set no other constraint on this foreground component, i.e. we assume no particular parametrization of it.

\subsection{Separation strategy}

We propose to extract the foreground component by subtracting an
estimate of the CMB from the total maps, and then post-processing the
remaining set of CMB--cleaned WMAP data to get maps of foreground
emission.  The effectiveness of this procedure is discussed in section
\ref{sec:discussion} and appendix~\ref{app:multi-ilc}.

Considering the minimal assumptions made about the foregrounds, it is
quite natural to use the so-called Internal Linear Combination method
(ILC) to extract a CMB map, to be subtracted from WMAP data to produce
CMB-free foreground maps. The ILC is a multifrequency linear filter
which uses a linear combination of the input data. It minimizes the
variance of the reconstructed CMB map over domains of observation as,
e.g, regions of the sky, or domains in harmonic space, or domains in
needlet space.

The strategy for obtaining foreground maps is thus the following:
\begin{itemize}
\item estimate a CMB map using an ILC;
\item subtract that CMB from the WMAP frequency maps (at proper resolution);
\item filter to maximize the signal to noise ratio in each of the foreground maps.
\end{itemize}

\subsection{The needlet ILC}

\begin{figure}
  \begin{center}
    \begin{tabular}{c}
      \includegraphics[width=50mm,angle=90]{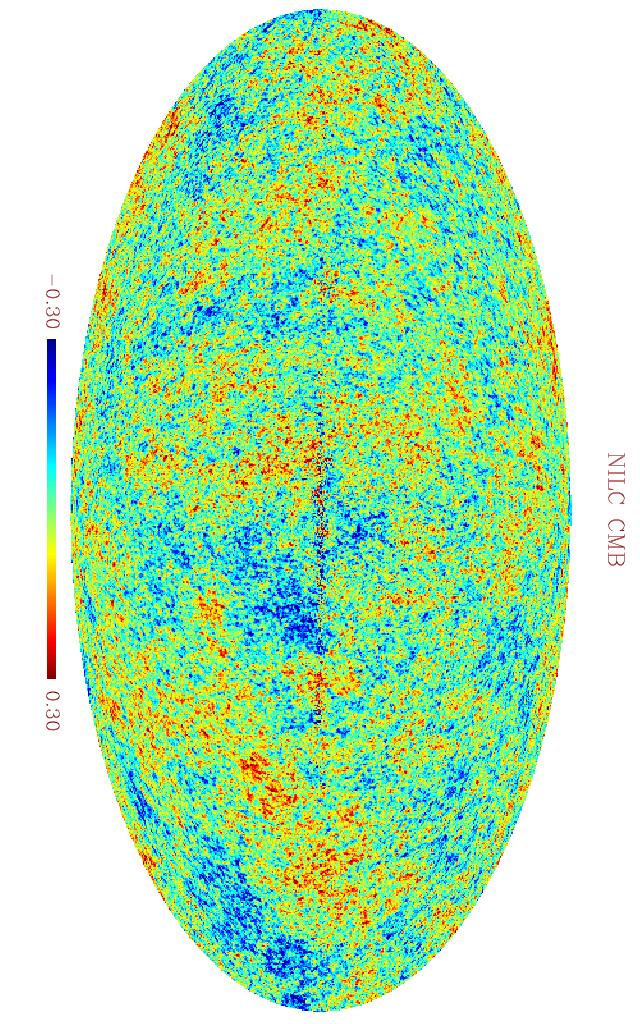}\\
    \end{tabular}
    \caption{The clean CMB map extracted using internal linear
      combination in needlet space, used for CMB subtraction in the
      present analysis}
    \label{fig:mapnilc}
  \end{center}
\end{figure}

Considering the variability of the statistical properties of foregrounds and noise both in pixel space and in harmonic space, the optimisation of the separation calls for localisation of the filter in both spaces. 
That can be achieved with needlets, which are a special type of wavelets on the sphere \citep{narcowich:petrushev:ward:2006,2008MNRAS.383..539M,guilloux:fay:cardoso:2008}. A needlet--based ILC is well suited to our component separation problem, as localisation in direct space (in addition to localisation in harmonic space) permits better reconstruction of the CMB in the vicinity of the galactic plane, i.e. a region of major interest for foreground science. 
For the present analysis, we use for CMB subtraction the WMAP 5-year needlet-based ILC map (NILC map) obtained by \cite{2009A&A...493..835D}, and displayed in figure \ref{fig:mapnilc}. Other options were considered, as for instance the harmonic ILC map derived in \citet{2006ApJ...645L..89S,2008PhRvD..78b3003S}, as has been done in \citet{2009PhRvD..79l3011G}. The NILC map has been chosen here because it reconstructs best the CMB in regions where foregrounds are significant.
It can be written as:
\begin{equation}
  \label{eq:hatslm}
  \hat s_\lm = b_\ell^W s_\lm + \delta_\lm
\end{equation}
where $\delta_\lm$ is the CMB reconstruction error and where the
beam is denoted $b_\ell^W$, corresponding to the resolution of channel~W.
The angular spectrum of that map is 
\begin{equation}
  \label{eq:asnilc}
  \mathrm{E} |\hat s_\lm|^2 = B_\ell^W  C_\ell^\mathrm{CMB} + N_\ell^{\rm ILC} 
\end{equation}
with $N_\ell^{\rm ILC} = \mathrm E(\delta_\lm)^2$ and
$B_\ell^W=(b_l^W)^2$.

\subsection{Channel--dependent CMB subtraction}

In our present application, the objective is to subtract the CMB in
each channel, without adding significant additional noise (and,
obviously, without adding more noise than we subtract CMB).  In each
channel, the total emission observed (eq. \ref{eq:obs-alm}) comprises
a CMB contribution $b_\ell^is_\lm$.  
We subtract from each observation map $ x_\lm^i$ a rebeamed,
rescaled version of the needlet ILC map, that is, the foreground map
$f_\lm^i $ at the resolution of channel $i$ is estimated by
\begin{equation}\label{eq:getfirstmap}
  \hat f_\lm^i = x_\lm^i - \alpha_\ell \frac{b_{\ell}^i}{b_\ell^W} \hat s_\lm
\end{equation}
where the coefficient $\alpha_\ell$ can be chosen to minimise the
reconstruction error, denoted $\varepsilon_\lm^i$, and defined
by:
\begin{equation}
  \label{eq:fi_chapeau}
  \hat f_\lm^i = b_l^i f_\lm^i + \varepsilon_\lm^i .
\end{equation}
Indeed, eqs~(\ref{eq:obs-alm}),~(\ref{eq:hatslm})
and~(\ref{eq:getfirstmap}) show that
\begin{equation}
  \label{eq:decepslmi}
  \varepsilon_\lm^i
  =
  (1-\alpha_\ell) b_\ell^i s_\lm 
  - \alpha_\ell \frac{b_{\ell}^i}{b_\ell^W} \delta_\lm
  + n^i_\lm 
\end{equation}
Neglecting the correlation between $\delta_\lm$ and $n_\lm^i$, one
gets
\begin{equation}\label{eq:msqerrec}
  \mathrm E ( \varepsilon_\lm^i )^2
  =
  (1-\alpha_\ell)^2 B_\ell^i C_\ell^{\rm CMB} 
  + \alpha_\ell^2 \frac{B_{\ell}^i}{B_\ell^W} N_\ell^{\rm ILC}
  + \mathrm E ( n_\lm^i )^2
\end{equation}
where $B_\ell^i = (b_\ell^i)^2$.  The mean square error $ \mathrm
E(\varepsilon_\lm^i )^2$ is easily found to be minimal for
\begin{equation}\label{eq:bestalphaell}
  \alpha_\ell = \frac{C_\ell^{\rm CMB}}{C_\ell^{\rm CMB} + N_\ell^{\rm ILC} / B_\ell^W}.
\end{equation}
This is exactly the signal to signal-plus-noise ratio
in~(\ref{eq:asnilc}), that is, the prescription for Wiener filtering.
Therefore, the variance of the foreground map in channel $i$ (at the
resolution of the channel), is minimal if we subtract the
Wiener-filtered version of the needlet ILC map.

\begin{figure*}
\begin{center}
\begin{tabular}{cc}
\includegraphics[width=50mm]{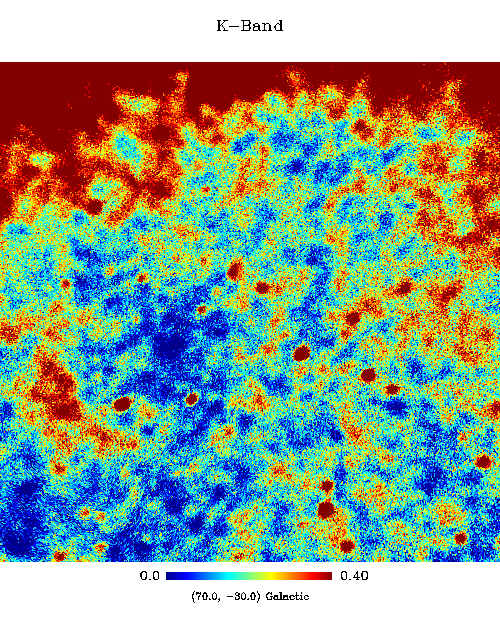}&
\includegraphics[width=50mm]{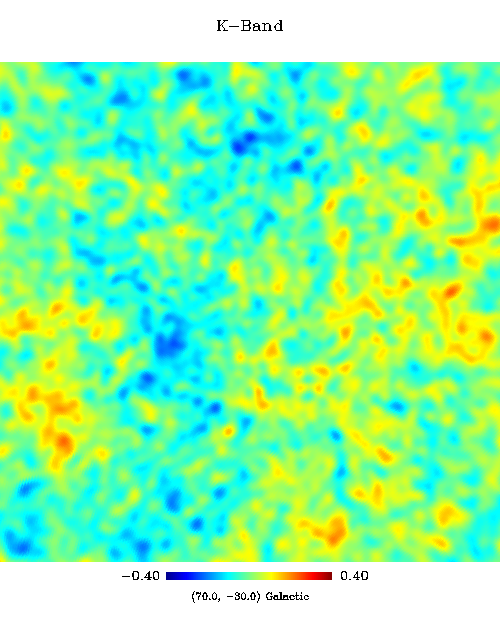}\\
\includegraphics[width=50mm]{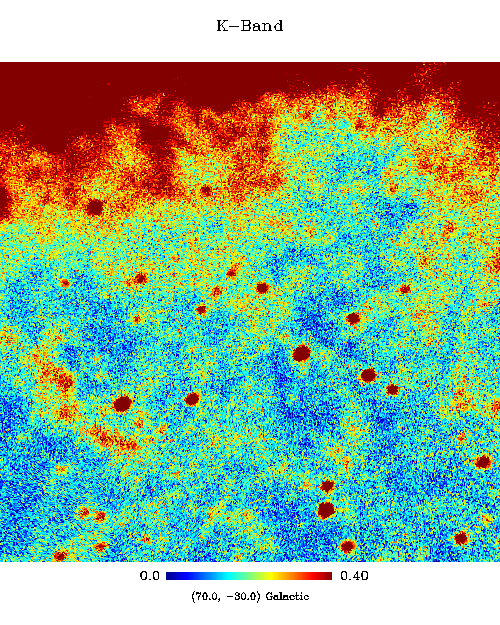}&
\includegraphics[width=50mm]{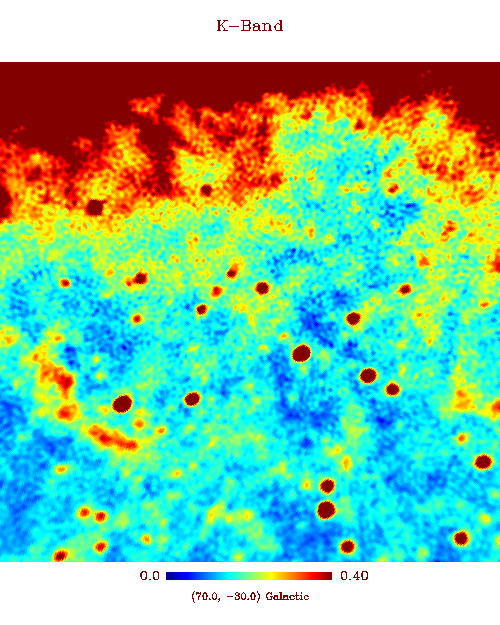}\\
\end{tabular}
\caption{(a - top left) The original K band map, (b - top right) The needlet CMB map at K band resolution, (c - bottom left) Map after CMB subtraction, (d - bottom right) Filtered foreground map. The improvement from a) to c), and then from c) to d), is striking.}
\label{fig:demo}
\end{center}
\end{figure*}

CMB subtraction according to~(\ref{eq:getfirstmap}) yields our primary
foreground map in WMAP channel $i$, at the resolution of the
considered WMAP channel.  The map contains essentially foreground
emission and noise from channel $i$, and additional error originating
from imperfect CMB subtraction.  This additional error is discussed in
more detail in section \ref{sec:noise}.
Figure \ref{fig:demo} illustrates for one of the WMAP channels (here,
the K band at 23 GHz), on a patch centered at modest galactic latitude
(latitude -30 degree and longitude 70 degree), the separation of the
original observation into CMB and foregrounds.  The effect of CMB
subtraction is clearly visible when comparing the original map (top
left) to the map after CMB has been subtracted (bottom left).
Most of the large to intermediate scale fluctuations due to the
presence of CMB have been removed, which permits to identify clearly
compact sources and structures of the galactic emission.

\subsection{Latitude-dependent filtering} \label{sec:wiener}

After CMB subtraction, the estimated foreground map $\hat f_{\ell
  m}^i$ still contains significant noise, which dominates on small
scales.  This is clearly visible in the bottom left panel of figure
\ref{fig:demo}.  That map is obtained, by construction, at the
resolution of the original input map but we still have extra degrees
of freedom: adjusting the beam or, equivalently, multiplying the
harmonic modes by some filter $w_\ell^i$, one obtains a new estimate
$\tilde f_\lm^i$ of the foreground map as:
\begin{displaymath}
  \tilde f_\lm^i = w_\ell^i \hat f_\lm^i 
\end{displaymath}
The filter can be adjusted to minimize the total variance
\begin{displaymath}
  \sum_{\ell} \sum_{m=-\ell}^\ell 
  \mathrm E \left | \tilde f_\lm^i -  b_\ell^i f_\lm^i \right |^2,
\end{displaymath}
and it is well known that, in the ideal case where both the foreground
map of interest $f_\lm^i$ and the contamination term
$\varepsilon_\lm^i$ are stationary, the best filter is the Wiener
filter:
\begin{displaymath}
  w_\ell^i = \frac{B_l^i C_\ell^{\rm FG}}{B_l^iC_\ell^{\rm FG} + E_\ell}  
\end{displaymath}
where $E_\ell$ is the variance of the reconstruction error $\varepsilon_\lm^i$.

However, we are not facing a stationary situation.
As galactic foregrounds are strongly concentrated in the galactic
plane, the filtering should better depend on the location on the sky.
It is not immediately clear what would be the best strategy for such a
localized filtering, especially if there is need not only to provide
the best map but also to characterize it.
As a reasonable compromise between efficiency and simplicity, we
approximate the foreground emission as following a plane parallel slab
model.  We then apply a filter for each zone of galactic latitude. 
As discussed in the next section, we also constrain the total response 
after filtering to have a Gaussian shape, i.e. we find the Gaussian beam 
which fits best the harmonic response $w_\ell^ib_\ell^i$ resulting from the
instrumental beam \emph{and} the Wiener filter.  
In that way, the variation of the beam across the sky can be described
simply by a latitude dependent (actually: depending only on the
latitude zone) Gaussian beam.  The improvement of the quality of the 
derived foreground map after filtering is clearly visible in figure
\ref{fig:demo}, where the bottom right panel displays the foreground 
map after the filtering. 
Details of the implementation 
are given in the next section.

\begin{figure*}
  \begin{center}
    \begin{tabular}{cc}
      \includegraphics[width=80mm]{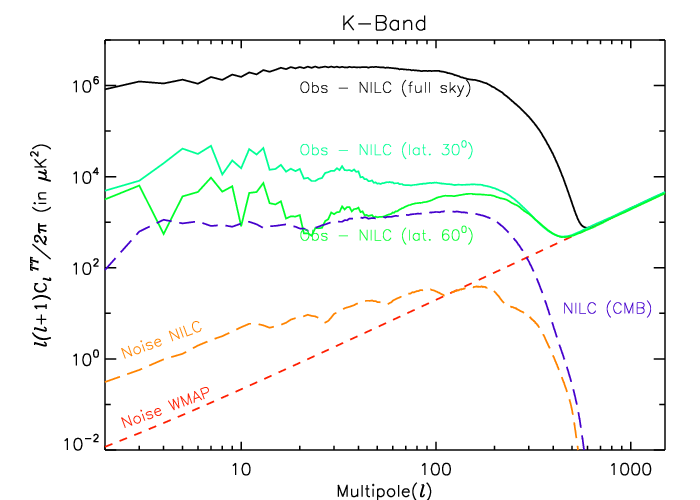}&
      \includegraphics[width=80mm]{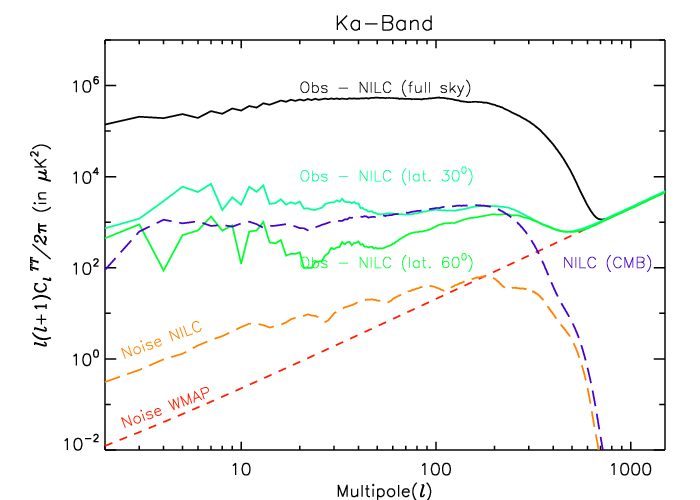}\\
      \includegraphics[width=80mm]{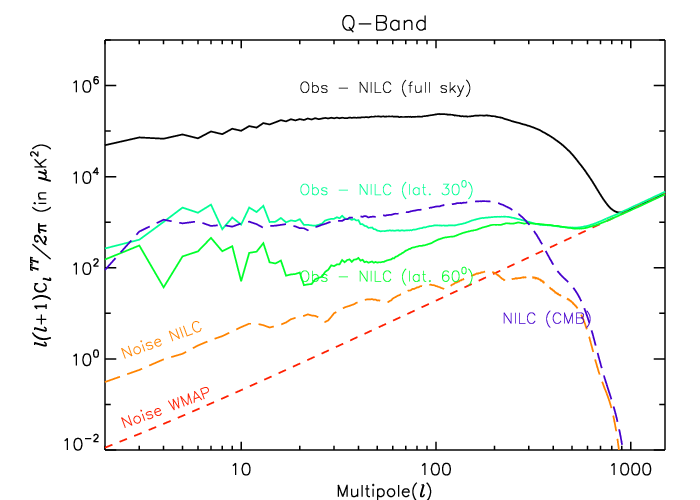}&
      \includegraphics[width=80mm]{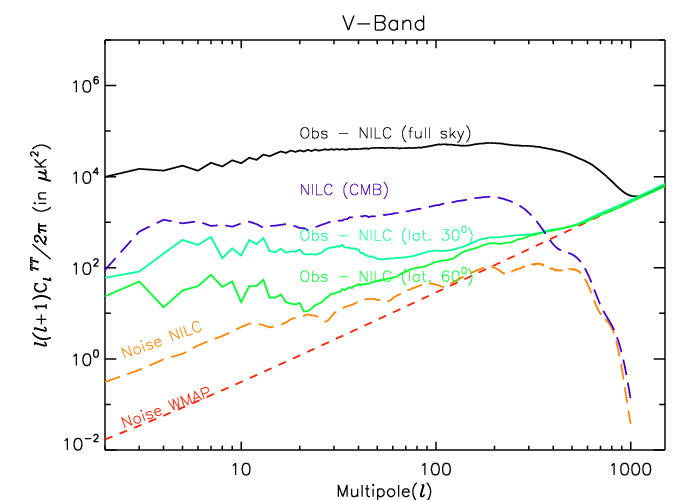}
    \end{tabular}
    \caption{Angular power spectra over the full sky for the frequency
      bands K to V. The smooth black curve is the power spectrum of the
      WMAP CMB-subtracted map, the red line (dashed) is the
      theoretical WMAP noise spectrum for the channel considered, and the orange
      line (double dashed) is the noise contribution from the NILC map. 
      Foreground (galactic) emission is dominant on large scale, whereas the theoretical WMAP noise nearly
      matches the power spectrum of the reconstructed map on small scale. The
      NILC noise lies somewhere between the 
      level of the CMB and that of the WMAP channel noise on large scales, and is significantly
      below the instrumental noise on small scales.}
    \label{fig:spectrum_bef}
  \end{center}
\end{figure*}

\begin{figure*}
  \begin{center}
    \begin{tabular}{cc}
      \includegraphics[width=80mm]{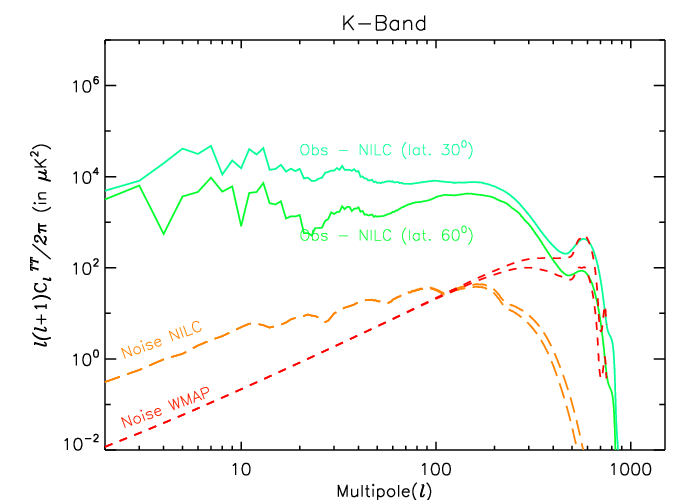}&
      \includegraphics[width=80mm]{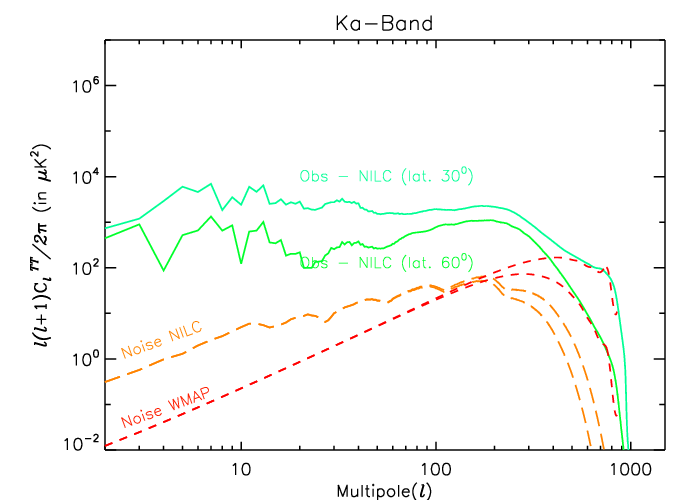}\\
      \includegraphics[width=80mm]{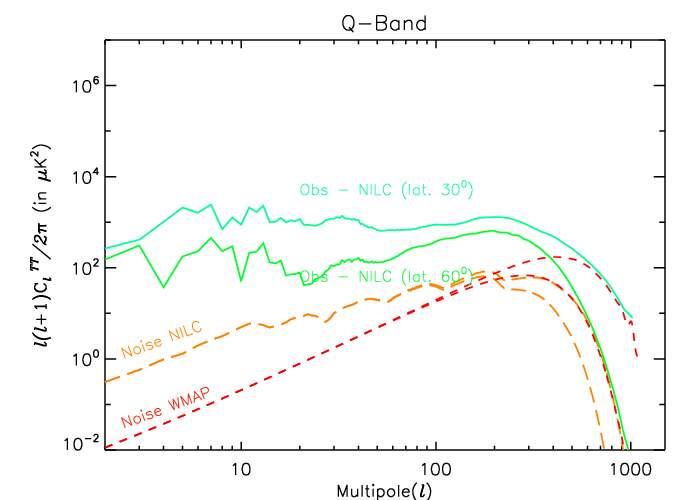}&
      \includegraphics[width=80mm]{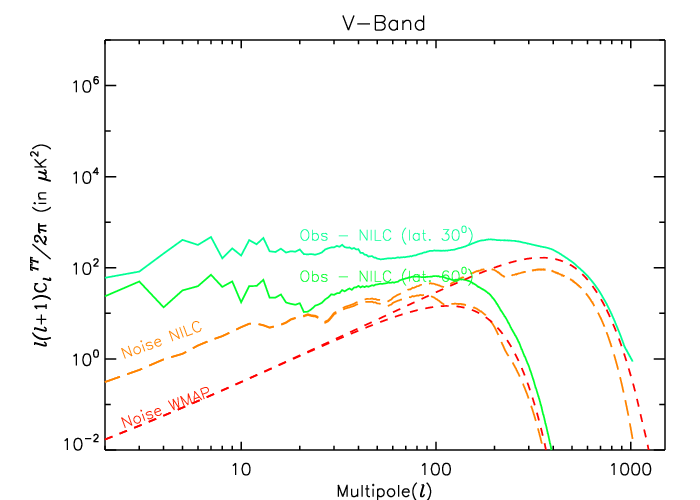}
    \end{tabular}
    \caption{Angular power spectra after the Wiener filtering for the
      frequency bands K to V. This is the equivalent of fig. \ref{fig:spectrum_bef} after noise filtering.
      The bump visible on small scales for channel K (and, to lesser extent, channel Ka), is due to the non--gaussian
      shape of the WMAP beams. It could be gotten rid of by relaxing the constraint that our final map should be characterised by a Gaussian beam.}
    \label{fig:spectrum_aft}
  \end{center}
\end{figure*}

\section{Practical implementation}

\subsection{CMB subtraction}

For each WMAP channel, we start from the Wiener filtered Needlet ILC
CMB from \cite{2009A&A...493..835D} which includes the $\alpha_\ell$
ratio of eq. \ref{eq:bestalphaell}.  The resolution of the map is changed in harmonic space, by
multiplication of each $a_\lm$ by the ratio $b_\ell^i/b_\ell^W$.  The
resulting CMB map is subtracted from the observation in channel $i$.
Note that because of the Wiener filter applied on the CMB map, the
equivalent resolution of this CMB map is \emph{not} that of channel
$i$.

The foreground map obtained in a given channel by subtraction of the
CMB contains foregrounds, noise, and a residual (the difference
between the CMB estimate and the true CMB at the resolution of the
given channel).  The residual contains some noise from all channels,
some residual CMB, and a small amount of foregrounds.  This residual,
however, is well below the foreground emission on large scales and
orders of magnitude below the noise of the considered WMAP band on
small scales.  This is illustrated, for four of the WMAP channels, in figure \ref{fig:spectrum_bef},
which displays the power spectra of the
WMAP observations after CMB subtraction, as compared to the original
CMB contamination, to contamination by noise, and to the residual due
to improper CMB subtraction.  
In all channels, the process of CMB subtraction reduces by two orders
of magnitude the total contamination power on the largest scales.  For
the V channel, a significant reduction of the total contamination is
achieved up to $\ell$ larger than 500.

\subsection{Latitude zones}
\label{sec:latzones}

The sky is divided into 7 latitude zones, selected using multiplicative masks. Each mask is defined as follows:
\begin{align*}
M(\theta,\phi) &= 1  \ \  &\text{for} \ \theta_0 < |\theta| < \theta_1\\
&=\cos^2 \left( \frac{\pi}{2} \times \frac{\theta_0-|\theta|}{\Delta \theta} \right) \ \  &\text{for} \ \theta_0-{\Delta \theta} \leq |\theta| < \theta_0 \\
&=\cos^2 \left( \frac{\pi}{2} \times \frac{|\theta|-\theta_1}{\Delta \theta} \right) \ \  &\text{for} \ \theta_1 < |\theta| < \theta_1+\Delta \theta \\
&=0 \ \ &\text{elsewhere.}
\end{align*}
Here $\Delta \theta$ is the width of transition regions of the mask
(regions in which its value decreases smoothly from 1 to 0). The
latitudes $\theta_0$ and $\theta_1$ are the lower and upper limits of
the region where the mask value is 1. Note that each region is
symmetric with respect to the galactic plane. Smooth transitions in
the masks for the different latitude bands simply permit to avoid edge
effects in the masked maps, mitigating mode mixing in the calculation
of power spectra.  Limits for the zones (i.e. values for $\theta_0$
and $\theta_1$) used in this analysis are 0, 5, 10, 20, 30, 45, 60 and
90 degrees, with transitions of $\Delta \theta = 5^\circ$. The width
of the regions selected in this way increases with increasing galactic
latitude, which is required for accurate estimation of the Wiener
filter (see below).

\subsection{Latitude-dependent Wiener filter}

For each channel, for each latitude band, we compute the power
spectrum $C_\ell^{\rm OUT}$ of the masked noisy foreground map.  For
each $\ell$, the total power comprises a contribution $F_\ell$ from
the foregrounds $b_\ell^i f_\lm^i$ at the resolution of channel
$i$ (i.e. the signal of interest) and a contribution $E_\ell$ from the
reconstruction error $\varepsilon_\lm^i$. 
In this case, the Wiener filter can be approximated as:
\begin{equation}
  \label{eq:practical-wiener}
  w_\ell = \frac{C_\ell^{\rm OUT} - E_\ell}{C_\ell^{\rm OUT}}.
\end{equation}
See section~\ref{sec:noise} for the estimation of $E_\ell$.  The
Wiener filter as defined by equation \ref{eq:practical-wiener} is
applied independently in each WMAP channel.

The effective beam after Wiener filtering becomes $w_\ell b_\ell^i$.
That is the beam which minimizes the variance of the error of the
reconstructed foreground map, for channel $i$, in the latitude band
defined by the mask corresponding to the zone considered.  In fact,
such a Wiener filter can be computed locally for any region of
interest, not only for latitude bands, which are used here mainly for
illustrative purposes.

\subsection{Gaussian beam approximation}

As discussed above, it is not very convenient to construct a final
foreground map in which the beam varies rapidly over the sky.  For
simplicity, it may also be desirable to work with effective Gaussian
beams, in which case the beam is fully defined by a single number: its
full width at half maximum (FWHM).  The beam response in harmonic
space then is:
\begin{equation*}
  b_\ell^{\rm gauss} = \text{exp}\left(-\frac{\ell(\ell+1)\sigma^2}{2}\right)
\end{equation*}
where $\sigma$ is related to the FWHM of the Gaussian beam by the
relation $\sigma = \theta_{\rm FWHM}/ \sqrt{(8\times \log 2)}$.

\begin{figure}
\begin{center}
\includegraphics[width=85mm]{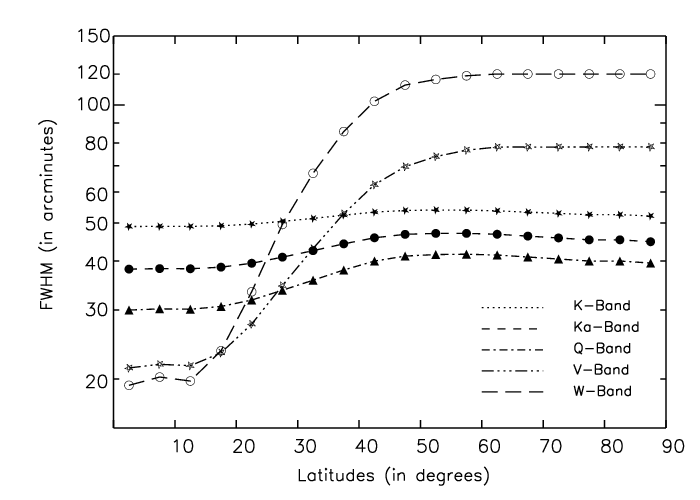} 
\caption{The FWHM of the Gaussian beam used to approximate the Wiener filter, as a function of latitude, for all WMAP channels. For each WMAP frequency channel, the expected general behaviour is observed: the resolution can be better at low galactic latitude, where there is more signal, and hence better signal to noise ratio at high $\ell$.}
\label{fig:beam}
\end{center}
\end{figure}

For each of our seven selected regions then, we determine the Gaussian
beam which best matches the theoretical optimal effective beam due to
the effect of both the instrumental beam and the Wiener filter. 
We derive a FWHM which varies smoothly in latitude bands of 5 degree
width by interpolating the FWHM derived using the broader latitude
bands described in \ref{sec:latzones} above.  The corresponding FWHM,
for all WMAP channels and all latitudes in 5 degree steps, is plotted
in figure \ref{fig:beam}.

Note that while our choice results from a trade-off between optimality
and convenience, it is somewhat arbitrary. We could, for instance,
have decided to work with more complex symmetrical beam shapes or
different zones.  We stress, however, that no information is lost in
that process since one can always change back the response in harmonic
space (and change the effective beam) by multiplication of $a_\lm$
coefficients by the appropriate response ratio.  The FWHM plotted in
figure \ref{fig:beam} are just indicative of the scales where noise
starts to dominate over the foregrounds, as a function of channel and
galactic latitude.

\subsection{Construction of the filtered maps}
\label{sec:practical-filtering}

After having determined, for each frequency channel, the resolution at
which the foreground map should be reconstructed as a function of
galactic latitude, a filtered final map for each WMAP channel is
obtained in the following way.  
Starting from the original WMAP observations and after CMB removal, we
make 18 different maps, one at the appropriate resolution for a 5--degree wide
latitude band. The final map for that channel is obtained by picking,
for each pixel of the final map, the corresponding pixel in the one of
the 18 maps which is at the resolution matching the latitude band of
the pixel at hand. Hence, the resolution of the filtered map varies in
steps over the sky, with a change in resolution at all multiples of
$5^\circ$.

\section{Discussion}\label{sec:discussion}

\subsection{Multi-frequency analyses}

As seen in figure \ref{fig:beam}, the Gaussian beam which maximises
the signal to noise ratio of the foreground map in a given channel
depends both on galactic latitude and on the channel.

Further analysis of these foreground maps, however, involves measuring
spectral indices in selected regions or the spectral energy
distribution of compact sources, i.e. ratios between flux at different
frequencies. Such post-analyses require maps at the same resolution.

For specific, localised analyses, using several WMAP channels, original CMB-cleaned maps can easily be re-beamed for better matching of the resolution with the local S/N ratio. Figure \ref{fig:beam} provides a helpful tool for choosing the resolution of the analysis. To take a specific example, for investigations involving all channels at 25 degrees latitude, one should pick a resolution somewhere between 30 and 50 arc-minutes, as the curves giving the appropriate resolution, for each channel, in that latitude zone, give numbers between 30 and 50.

\begin{figure*}
\begin{center}
\begin{tabular}{cc}
\includegraphics[width=50mm]{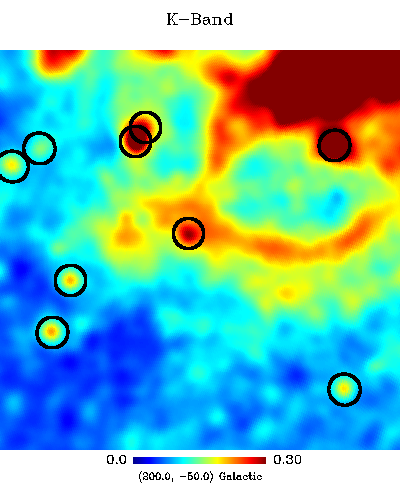}&
\includegraphics[width=50mm]{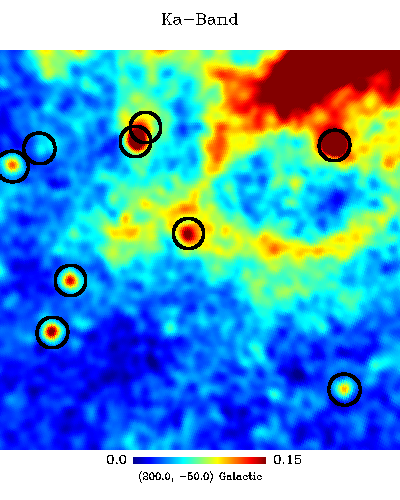}
\end{tabular}
\caption{40 degree by 25 degree patch of the sky at latitude of -50 and longitude of +200. The K and Ka band foreground maps are shown in this range. Most of the point sources (circled by black) identified by the WMAP team are visible by eye. }
\label{fig:clc}
\end{center}
\end{figure*}

\subsection{Faint foregrounds?}

Although the original claim in the present work is the construction of maps of all foregrounds, it should be noted that the final maps are appropriate only for the scientific analysis of foregrounds whose contribution to the observed emission is larger than the WMAP noise (but can be smaller than the CMB), i.e. of galactic foregrounds and strong point sources.

The reason is as follows. The subtracted CMB has been obtained by an ILC method, which finds locally linear combinations which minimise the total variance of the CMB reconstruction error. When subdominant foregrounds are present, it is more favourable to the ILC to let weak foregrounds leak in the CMB map and reduce the instrumental noise, than the opposite. For instance, Sunyaev Zel'dovich effects are too weak to play a role in the adjustment of the linear combinations used for CMB reconstruction in WMAP data. The same is true for a weak background of point sources. Hence, the ILC is not optimised to reject these foregrounds, which thus are not preserved in the final foreground maps produced with our processing. A search for SZ signals in WMAP data has been performed by \citet{2010arXiv1001.0871M} using a method appropriate for this particular signal, the multifrequency matched filter \citep{2002MNRAS.336.1057H,2006A&A...459..341M}.

\subsection{Noise levels}
\label{sec:noise}

The `noise' in our foreground maps (from all sources of error) in the map for channel $i$ (at the resolution defined by the beam $b_\ell^i$) comes from four main terms:
\begin{enumerate}
\item noise $n^i$ from the relevant WMAP channel;
\item noise $\sum w_j n^j$ from `internal linear combinations' of WMAP channels, present in the CMB map subtracted from each WMAP observation;
\item foregrounds leaking in the needlet ILC estimate of the CMB;
\item CMB residuals after subtraction of the estimated CMB.
\end{enumerate}
An exact characterisation of the total error is not completely
possible in the absence of a reliable model of the foregrounds, which
is needed to estimate the amount of foreground emission leaking in the
estimated CMB.  It is possible, however, to give an estimate of the
total `noise' power spectrum, $E_\ell$.

\subsubsection{Noise power spectrum estimate}

An estimate is obtained by considering that all terms are independent
so that the total power of the error is the sum of the powers of all
contributions.

The first term is simply the noise power $N_\ell^i$ of the WMAP $i^{\rm \it{th}}$ channel, the level of which is provided with the release of WMAP data.

The second term is estimated by Monte-Carlo simulations of the WMAP noise by 
\citet{2009A&A...493..835D}. Noise maps and power spectra released are available on a dedicated web 
page\footnote{\tiny http://www.apc.univ-paris7.fr/APC/Recherche/Adamis/cmb\_wmap-en.php}.
They should be corrected from the effect of the Wiener filtering and rebeaming. This is done by multiplying each mode, in harmonic space, by:
\begin{equation}
  r_\ell = \frac{b_\ell^i \, w_\ell}{(b_\ell^{\rm NILC})^2} 
\end{equation}
where $w_\ell$ is the effective beam of the Wiener-filtered needlet
ILC CMB map, $b_\ell^i$ is the beam of the foreground map, and
$b_\ell^{\rm NILC}$ the beam of the NILC map (coinciding, for $\ell <
1000$, with that of the W channel. The power $N_\ell^{\rm CMB}$ of the
error from this term is:
\begin{equation}
  N_\ell^{\rm CMB} = r_\ell^2 N_\ell^{\rm ILC}
\end{equation}
where $N_\ell^{\rm ILC}$ is the noise level, per harmonic mode, available with the needlet ILC CMB map.

The third term is unknown. On the vast majority of the sky where
foregrounds are weak, it is negligible.  Where foregrounds are strong,
it is always a small fraction of the input foregrounds.  It is
therefore neglected in our noise estimate.

The fourth term comprises contributions from two effects: 
\emph{a}) residual CMB from the difference between the beam of the WMAP channel considered and the effective beam of the subtracted CMB, which can be computed straightforwardly; \emph{b})
CMB modes cancelled by the ILC because of empirical correlation with the foregrounds and noise (see appendix and figure 6 and of \citet{2009A&A...493..835D}). 

Appendix \ref{app:recon-noise} gives a detailed calculation of all noise terms. 
The power $E_\ell$ of the sum of all contributions (total error) is given by the expression:
\begin{multline}
  E_\ell \simeq N_\ell^i + r_\ell^2 N_\ell^{\rm ILC} +  \\
  \left[ \left(b_\ell^i\left(1-\alpha_\ell\right)\right)^2 + 0.04 \,
    b_\ell^i\left(1-\alpha_\ell\right)b_\ell^W r_\ell
  \right]C_\ell^{\rm CMB}
  \label{eq:noise-El}
\end{multline}
where $C_\ell^{\rm CMB}$ is the CMB power spectrum (which can be taken
to be the WMAP best fit model, for instance). The last term is the
total contribution from CMB subtraction error. The first term in the
brackets comes from effect \emph{a}) above, and the second term is a
correction originating from the 2\% CMB power loss due to the ILC
bias.

\begin{figure}
\begin{center}
\includegraphics[width=85mm]{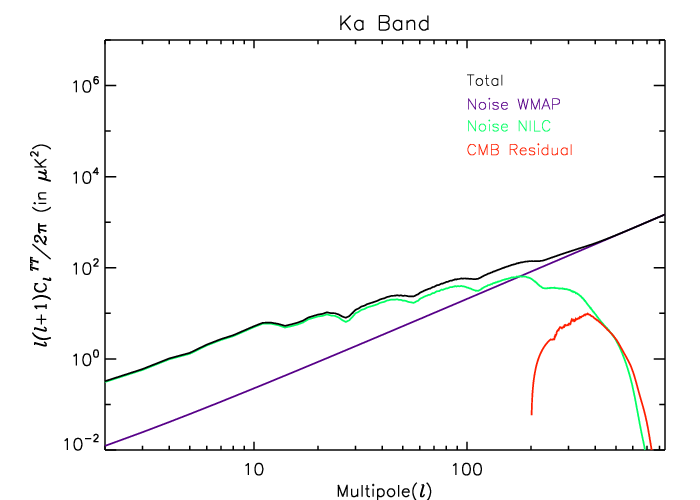}
\caption{Power spectra of the noise in the Ka band, decomposed into three main sources of error. The instrumental noise from that frequency channel dominates on small scales, whereas CMB reconstruction error (noise in the NILC map) dominates on large scales. The contamination of the foreground map by the CMB is very small.}
\label{fig:noisedecomp-Ka}
\end{center}
\end{figure}

Figure \ref{fig:noisedecomp-Ka} shows the decomposition of the total
noise into three main terms: noise from WMAP in channel Ka, noise from
the NILC map, and residual CMB improperly subtracted.

\begin{figure}
\begin{center}
\includegraphics[width=85mm]{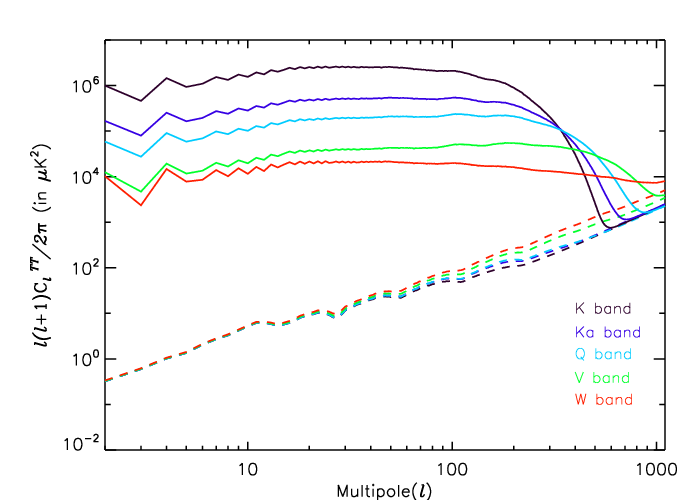}
\caption{Power spectra of foreground maps and corresponding noise level.}
\label{fig:signalandnoise}
\end{center}
\end{figure}

Figure \ref{fig:signalandnoise} displays the power spectra of all foreground maps, and power spectra of the noise term in all of them.

\subsubsection{Noise power spectrum upper limit}

The two first terms (i) and (ii) (arising from WMAP instrumental
noise) are correlated because the instrumental noise in the estimated
CMB contains a part of the noise on the channel of interest.  An upper
limit to the total error is obtained by assuming that the correlation
is perfect between $n^i$ and $\sum w_j n^j$, which is the case when
channel $i$ dominates with a negative coefficient in the ILC
(negative, because the ILC is subtracted from channel $i$ to get the
foreground map).

Hence, we can assume the following upper limit for $E_\ell$:
\begin{multline}
  E_\ell  \leq  
  \left[ 
    \left ( N_\ell^i \right )^{\half}  + 
    r_\ell  \left ( N_\ell^{\rm ILC} \right ) ^{\half} 
  \right] ^2 + \\
  \left[
    \left(b_\ell^i\left(1-\alpha_\ell\right)\right)^2 +
    0.04 \, b_\ell^i\left(1-\alpha_\ell\right)b_\ell^W r_\ell
  \right]C_\ell^{\rm CMB}
  \label{eq:upper-El}
\end{multline}

\subsubsection{Noise level maps}

The origin of all main contributions to noise being identified, it is a straightforward matter to write a pixel space equivalent of equations \ref{eq:noise-El} and \ref{eq:upper-El}.

\subsubsection{More accurate noise characterisation ?}

The description of noise as a single noise power spectrum per map only
is an approximation.  It is convenient for most applications, but one
should be aware of its limitations.

The first limitation comes from the fact that the noise is not
stationary for two reasons. The first reason is that the WMAP hit
count is not homogeneous on the sky. The second reason is that
galactic foregrounds, and hence the coefficients of the needlet space
ILC used to subtract the CMB, and hence the noise from the ILC map,
are non stationary.  If needed, it is possible to scale the two first
terms of equation \ref{eq:noise-El} as a function of the local noise
levels. One can also generalise equations \ref{eq:noise-El} and
\ref{eq:upper-El} using the pixel-based noise variance of WMAP channel
$i$ in place of $N_\ell^i$, and using noise simulations provided by
\citet{2009A&A...493..835D}, filtered by $r_\ell$, in place of
$N_\ell^{\rm ILC}$.

The second limitation comes from the cross-channel correlation of the
resulting noise.  All terms in the right hand side of
equation~\ref{eq:noise-El}, except the first term, give rise to such
correlated noise.  On large scales, these correlated terms dominate
the total error, so the noise is close to 100\% correlated. On small
scales, the noise from WMAP $i^{\rm \it{th}}$ channel dominates, so
that the noise is not correlated between channels.

More accurate noise characterisation are possible, but they require
Monte-Carlo simulations of the whole process.  If needed for very
precise analyses, such simulations (including in particular cross
correlation of the error between channels) can be performed by the
authors upon request.

\subsection{Products}

Our foreground maps and related data comprise one CMB-cleaned map
(i.e. a foreground map) for each WMAP channel, at the original
resolution of the channel, characterised by a corresponding noise
power spectrum (given by equation \ref{eq:noise-El}) for each channel
and an effective beam for each channel (copied from the beam provided
with the WMAP 5-year data).

For each channel, a suggested resolution (Gaussian full width at half
maximum) as a function of galactic latitude, appropriate for
minimizing the foreground reconstruction error at that latitude, is
given in figure~\ref{fig:beam}.

\begin{figure}
\begin{center}
\begin{tabular}{cc}
\includegraphics[width=40mm,angle=0]{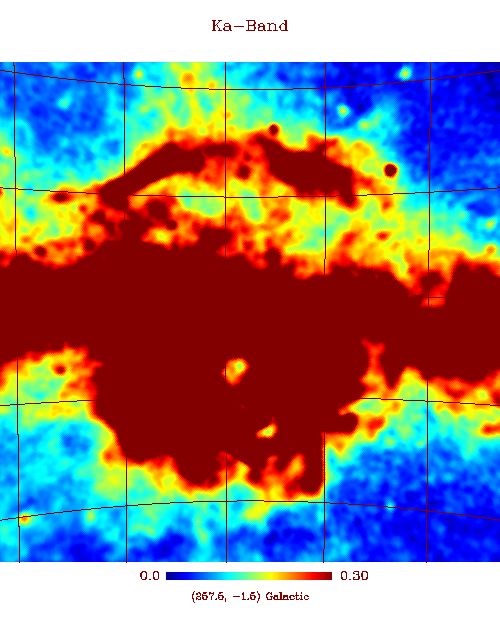}&
\includegraphics[width=40mm,angle=0]{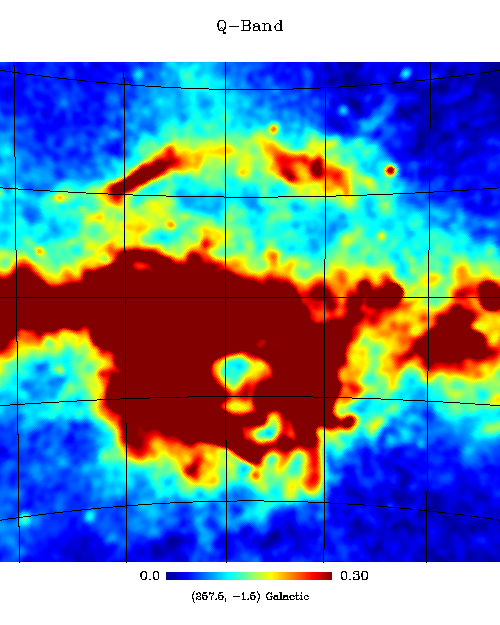}\\
\includegraphics[width=40mm,angle=0]{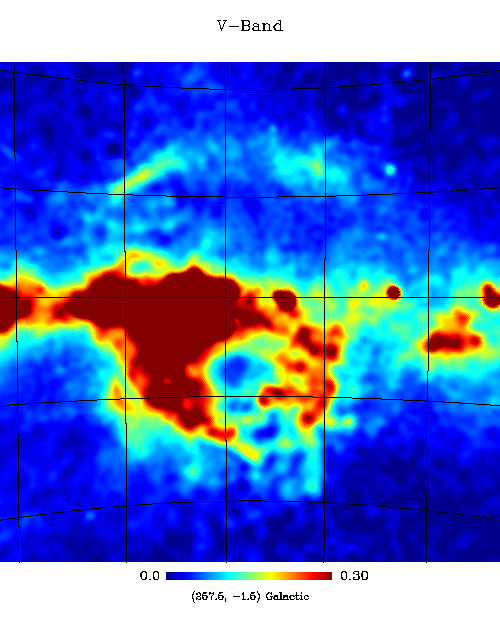}&
\includegraphics[width=40mm,angle=0]{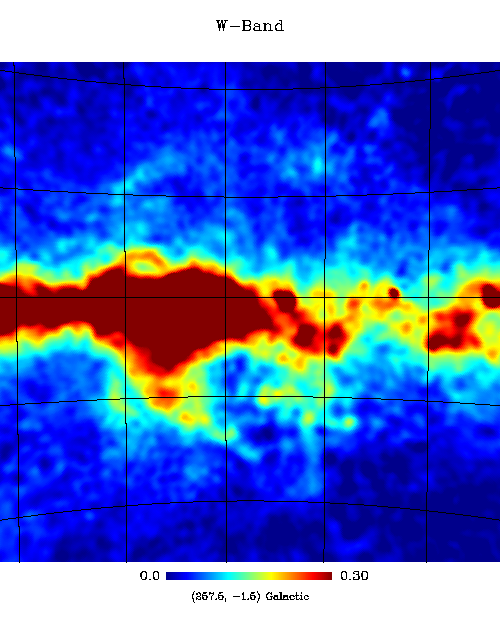}\\
\end{tabular}
\caption{Gum Nebula as viewed from Ka to W band at K-band resolution. K band resolution is used to show the diffuse emission rather than the small scale features.}
\label{fig:gumnebula}
\end{center}
\end{figure}

Figure \ref{fig:gumnebula} displays our filtered foreground map for channels Ka to W in the region of the Gum nebula.
Figures \ref{fig:fullskymaps1} to \ref{fig:fullskymaps5} display full sky foreground products, both at the original WMAP resolution, and with noise filtered out (latitude dependent beams). Contribution from different astrophysical processes can be seen even in the high frequency channels on the filtered maps.

\subsection{Comments about the present approach}

The approach to foreground estimation discussed in this paper is quite
possibly the simplest way to achieve reliable foreground estimation.

The foreground products delivered by the procedures described in this
paper are obtained with little prior information: the input maps are
supposed to be well calibrated and they are supposed to be accurately
characterized by their beams and noise properties.  Besides that, no
prior information is used about the foreground emissions, and no
attempt is made at constraining or modeling those emissions.  As a
result, we obtain `maximal' foregrounds maps.

It would be interesting to aim at producing more constrained
foreground maps by including prior information about foreground
emission.  In the SMICA approach, for instance, the foreground
emission is typically modeled as the superposition of a number $\rfg$
correlated templates.  This, however, is far from being as
straightforward as the ILC-subtraction proposed in this paper.  

It should be noted that the ILC-subtraction approach based on $\nobs$
observation frequencies can be shown to be equivalent to assuming that
the foreground emission can be represented as the superposition of
$\rfg=\nobs-1$ templates (see appendix for a proof and a more careful
statement).  Along the same lines, one may also consider a
multi-dimensional version of the ILC targeted directly at the
foregrounds.  This is briefly discussed in the appendix and is the
topic of a forthcoming publication \cite{Mathieu}.

\subsection{Comparison with other WMAP foreground maps}

Other maps of WMAP galactic foregrounds have been obtained by a variety of methods.
All such maps, however, are produced at resolution of 1 degree or worse. The maps obtained in the present work are provided at the resolution of the original WMAP channels.

In addition, as the method subtracts a CMB map significantly cleaner, in the galactic plane, than that obtained by a simple ILC, our maps give better estimates of the total emission of the ISM in the vicinity of the galactic plane.

\section{Conclusion}

In the present paper, we have analysed the WMAP 5-year temperature maps
to clean them from the emission originating from the cosmic microwave background.
The maps obtained are noisy foreground maps, where the astrophysical emission is
dominated by emission from the galactic interstellar medium, and a small number of
compact sources.

The noise in these maps can be reduced by filtering. We estimate, as a
function of galactic latitude, the Gaussian beam to be used for
maximising the contrast of the signal of interest in the map.

The maps produced should not be used for searching for faint components such as the Sunyaev Zel'dovich effect or the emission from a background of faint sources. They are adequate for studying the emission of the galactic interstellar medium and the spectral energy distribution of strong compact sources.
Products can be downloaded from a dedicated web page.\footnote{\tiny {http://www.apc.univ-paris7.fr/APC/Recherche/Adamis/fg\_wmap-en.php}}

\section*{Acknowledgements}

Tuhin Ghosh thanks Indo-French Centre for the Promotion of Advanced
Research (IFCPAR) for the financial support for one month visit to
Paris during which this work was initiated. Some of the results in
this paper have used the HEALPix Package \citep{gorski05}. We acknowledge the use
of the Legacy Archive for Microwave Background Data Analysis
(LAMBDA)\footnote{\tiny {http://lambda.gsfc.nasa.gov/}}. Support for LAMBDA is provided by the NASA office of Space
Science.

\appendix
\label{sec:appendix}
\onecolumn

\section{ILC and ILC subtraction as oblique projections}
\label{app:multi-ilc}

\def\adj{^{t}}
\def\inv{^{-1}}

\def\va{\boldsymbol{a}}
\def\vx{\boldsymbol{x}}
\def\vf{\boldsymbol{f}}
\def\vg{\boldsymbol{g}}
\def\vw{\boldsymbol{w}}
\def\vz{\boldsymbol{z}}

\def\mB{\boldsymbol{B}}
\def\mF{\boldsymbol{F}}
\def\mG{\boldsymbol{G}}
\def\mI{\boldsymbol{I}}
\def\mP{\boldsymbol{P}}
\def\mR{\boldsymbol{R}}
\def\mU{\boldsymbol{U}}
\def\mW{\boldsymbol{W}}

In this appendix, we exhibit the direct connection between two
strategies for estimating the galactic emission: 
i) the strategy developed in this paper where the ILC-based estimate
is subtracted from the observations and ii) a direct estimation of the
galactic emission based on a multi-dimensional ILC.

That issue is discussed in the following context.
Since the processing takes place in harmonic space, the discussion can
be simplified by focusing on a single $(\ell, m)$ mode.  
Given then $\nobs$ input channels, denote $\vx$ the $\nobs\times 1$ vector of
the spherical harmonic coefficients of all channels after beam
correction at a particular $(\ell,m)$.
We can write $\vx=\va s+\vf$ where $s$ is the spherical harmonic
coefficient of the CMB for that mode and the $\nobs\times 1$ vector $\va$
contains channel gain with respect to CMB (it is a vector of $1$'s for
a perfectly calibrated instrument and input maps in the appropriate
units).  Vector $\vf$ represents all the other emissions.

\subsection{Component estimates as projections}

The ILC estimate of the CMB signal $s$ is $\hat s = \vw^t \vx$ where the
$\nobs\times 1$ vector $\vw$ is such that $\vw^t \vx$ has minimum variance
and has unit gain towards the CMB.  Hence, the ILC filter is the
minimizer of $E (\vw^t \vx)^2$ subject to $\vw^t \va=1$.  It is easily found
that the solution is $\vw^t =(\va\adj \mR\inv \va)\inv {\va\adj \mR\inv}$
where $\mR$ denotes the covariance matrix of $\vx$.  Hence, the CMB
reconstruction on all channels is the vector $\va\hat s = \mP_1 \vx$
where matrix $\mP_1$ is defined by
\begin{displaymath}
  \mP_1 = \frac{\va\va\adj \mR\inv}{\va\adj \mR\inv \va} .
\end{displaymath}
Matrix $\mP_1$ is a projection matrix ($\mP_1^2 = \mP_1$) but it is not an
\emph{orthogonal} projection ($\mP_1\neq \mP_1\adj$).  It is an oblique
projection onto $\mathrm{Span}(\va)$ along its null space $\mathcal{N}$
which, by definition, is the $(\nobs-1)$-dimensional subspace of $R^{\nobs}$:
\begin{displaymath}
  \mathcal{N} = \left\{ \vz \ |\  \va\adj \mR\inv \vz = 0 \right\} .
\end{displaymath}
This is the subspace of all directions which are nulled out by ILC.

In this paper, we considered reconstructing the foreground emission by
subtraction: $\hat \vf = \vx - \va \hat s$ so that the foreground estimate
is related to the data by $\hat \vf = \mP_2 \vx $ with
\begin{displaymath}
  \mP_2 = \mI_n - \mP_1 .
\end{displaymath}
By construction, $\mP_2$ is the projection matrix onto $\mathcal{N}$
along $\mathrm{Span}(\va)$.

Hence, the decomposition of the observed vector into CMB and
foregrounds corresponds to oblique projections onto a pair
$(\mathrm{Span}(\va), \mathcal{N})$ of complementary but non
orthogonal subspaces of dimensions~$1$ and $\nobs-1$ respectively.

\subsection{Multi-dimensional ILC}

One may consider a direct estimation of the foregrounds by
generalizing the ILC method to address the case of a multi-dimensional
component.  Consider then a $\rfg$-dimensional foreground model, that
is, the foreground vector is modeled as $\vf=\mF\vg$ where $\mF$ is a
fixed $\nobs\times \rfg$ full column rank matrix and $\vg$ is a
$\rfg\times 1$ random vector.  Then, just as in standard
(one-dimensional) ILC, one may estimate $\vg$ as $\hat \vg =
\mW^{\boldsymbol{\rm ILC}}\vx$ where $\mW^{\boldsymbol{\rm ILC}}$ is
an $\rfg\times \nobs$ matrix designed such that $\hat \vg$ has minimum
power under the constraint of offering unit gain to the foreground.
In other words, $\mW^{\boldsymbol{\rm ILC}}$ is the minimizer of $E
|\mW^{\boldsymbol{\rm ILC}} \vx|^2$ under the constraint
$\mW^{\boldsymbol{\rm ILC}}\mF=\mI_{\rfg}$.  The foreground emission
is then reconstructed as $\hat \vf= \mF \hat \vg = \mF
\mW^{\boldsymbol{\rm ILC}} \vx$.  Using the Lagrange multiplier
technique, it is readily found that
\begin{displaymath}
  \hat \vf = \mP_f \vx 
  \qquad\text{where}\qquad
  \mP_f = \mF(\mF^t \mR \inv \mF)\inv \mF^t \mR\inv .
\end{displaymath}
Matrix $\mP_f$ is recognized as an oblique projection onto
$\mathrm{Span}(\mF)$ along the $\nobs-\rfg$ dimensional subspace
$\mathcal{N}_f$:
\begin{displaymath}
  \mathcal{N}_f = \left\{ \vz \ |\  \mF\adj \mR\inv \vz = 0 \right\}  .
\end{displaymath}
Hence, multi-dimensional ILC appears as a direct generalization of the
one-dimensional case.

It is interesting to determine the conditions such that $\mP_f=\mP_2$
because then the foreground estimation procedure by the CMB-ILC
subtraction could be understood as a minimum variance procedure (the
$\rfg$-dimensional ILC) \emph{and} the minimum variance result would be
obtained \emph{without even knowing} matrix $\mF$.
This is examined in the next subsection.

\subsection{Subspace estimation}
\label{sec:foregr-subsp-estim}

Under which conditions do we have $\mP_2=\mP_f$?  First, we should
have identical dimensions, \textit{i.e.} $\rfg=\nobs-1$.
Let us then consider the noise-free model, $\vx=\va s+\mF\vg$ where
$\mF$ is $\nobs\times(\nobs-1)$ and $\mathrm{Cov}(\vg)=\mG$.  The covariance
matrix of the data is
\begin{equation}\label{eq:caat}
  \mR = \mathrm{Cov}(\vx) = C_\ell^{\rm CMB}\, \va\va\adj + \mF\mG\mF\adj .
\end{equation}
Given $\mR$ and $\va$, Eq.~(\ref{eq:caat}) uniquely determines the
range of $\mF$ (see proof at the end of this section).
Further, the corresponding value of $\mP_f$ (which depends on $\mF$
only through its range) is precisely equal to projection matrix
$\mP_2$.  Therefore, we can conclude that the CMB-ILC subtraction
method is identical to the multi-dimensional foreground-ILC when the
galactic subspace (the range of $\mF$) is determined from
(\ref{eq:caat}).  Note that these conclusions still hold when
empirical values are used instead of ensemble averages.

\subsubsection{Proof}
We show how the range space of $\mF$ is determined from $\va$ and
$\mR=C_\ell^{\rm CMB}\va\va^t +\mF\mG\mF^t$ when $\mF$ has $\nobs-1$ columns.

Let $\mW$ be an $\nobs\times \nobs$ whitening matrix \textit{i.e.} satisfying $
\mW\mR\mW^t = \mI$ and denote whitened quantities with a tilde:
\begin{displaymath}
  \tilde \vx = \mW\vx ,\qquad 
  \tilde \va = \mW\va ,\qquad 
  \tilde \mF = \mW\mF .
\end{displaymath}
Consider the $\nobs\times \nobs$ matrix $\mU = [ \tilde \va \, (C_\ell^{\rm CMB})^{1/2} | \
\tilde \mF \mG^{1/2}]$.  Then $ \mU\mU^t = C_\ell^{\rm CMB}\, \tilde \va\tilde \va^t
+ \tilde \mF \mG \tilde \mF^t = \mW(C_\ell^{\rm CMB}\, \va\va^t +\mF\mG\mF^t)\mW^t =
\mW\mR\mW^t =\mI$.  Hence matrix $\mU$ has orthonormal columns so that
the range space of $\tilde \mF$ is orthogonal to vector $\tilde \va$.
Therefore, the range space of $\mF$ can be estimated as follows: 
take $\mW$ any square root of $\mR\inv$, 
compute $\tilde \va = \mW\va$, 
compute an $\nobs\times(\nobs-1)$ matrix $\mB$ whose columns are orthogonal to
$\tilde \va$ so that $\mathrm{range}(\mB)$ is an estimate for
$\mathrm{range}(\tilde \mF)$, 
finally compute an estimate of $\mathrm{range}(\mF)$ as
$\mathrm{range}(\mW\inv \mB)$.

\section{Power spectrum of the reconstruction error}
\label{app:recon-noise}

The temperature map at a given frequency channel of WMAP is given in
harmonic space by
\begin{displaymath}
  y_{\ell m}^i = b_\ell^i s_{\ell m}+b_\ell^i f^i_{\ell m}+n_{\ell m}^i, 
\end{displaymath}
where $i$ indexes the WMAP frequency bands, $b_\ell^i$ is the beam of
the considered WMAP channel, and $s_{\ell m}$, $f^i_{\ell m}$ and
$n_{\ell m}^i$ are respectively the CMB, the foregrounds and the
instrumental noise in the given frequency band.  The CMB-ILC estimate
at the W band resolution expands as
\begin{displaymath}
  \widehat{s}_{\ell m}   =  b_\ell^W s_{\ell m} + \delta_{\ell m},
\end{displaymath}
where the residual noise of ILC (difference input-output) is
\begin{displaymath}
  \delta_{\ell m}
  =
  \sum_i w_{ILC}^i 
  \left(b_\ell^W f^i_{\ell m} + {b_\ell^W\over b_\ell^i} n_{\ell m}^i\right).
\end{displaymath}
Here $w_{ILC}^i$ are the needlet--ILC weights. Our foreground estimate
is computed by subtracting the Wiener filtered CMB-ILC estimate at the
beam of the considered frequency channel:
\begin{displaymath}
  \widehat{f^i}_{\ell m} 
  =  y_{\ell m}^i - \alpha_\ell {b_\ell^i\over b_\ell^W} \widehat{s}_{\ell m}.
\end{displaymath}
The Wiener filter $\alpha_\ell$ has been applied to reduce the
residual noise of the CMB-ILC. It is thus given by
\begin{displaymath}
  \alpha_\ell  = \frac
  {(b_\ell^W)^2 C_\ell^{\rm CMB} }
  {(b_\ell^W)^2 C_\ell^{\rm CMB} + N_\ell^{\rm ILC}}
  ,
\end{displaymath}
where $N_\ell^{\rm ILC} = E\left[\left(\delta_{\ell m}\right)^2\right]$.
Therefore, the foregrounds estimate at the frequency channel $i$ is
\begin{displaymath}
  \widehat{f^i}_{\ell m}  =  b_\ell^i f_{\ell m}^i + \varepsilon^i_{\ell m},
\end{displaymath}
where the reconstruction error $\varepsilon^i_{\ell m}\equiv
\widehat{f^i}_{\ell m} - b_\ell^i f_{\ell m}^i$ expands as follows:
\begin{displaymath}
  \varepsilon^i_{\ell m}
  = 
  \left(1-\alpha_\ell\right) b_\ell^i s_{\ell m} - \alpha_\ell {b_\ell^i\over b_\ell^W} \delta_{\ell m}  +  n_{\ell m}^i.
\end{displaymath}
The noise power spectrum $E_\ell  =  E\left[\left(\varepsilon^i_{\ell m}\right)^2\right]$ at the frequency channel $i$ (diagonal term of the covariance matrix of the error) is the sum of several contributions. We have:
\begin{displaymath}
 E_\ell  =  \left[\left(1-\alpha_\ell\right) b_\ell^i\right]^2 E\left[s_{\ell m}^2\right] + \left({\alpha_\ell b_\ell^i\over b_\ell^W}\right)^2 E\left[\left(\delta_{\ell m}\right)^2\right] + E\left[\left(n_{\ell m}^i\right)^2\right]
           -2 \left(1-\alpha_\ell\right) b_\ell^i \left({\alpha_\ell b_\ell^i\over b_\ell^W}\right) E\left[ s_{\ell m} \delta_{\ell m}\right] -2 \left({\alpha_\ell b_\ell^i\over b_\ell^W}\right) E\left[ n_{\ell m}^i \delta_{\ell m}\right].
\end{displaymath}

We neglect the last term (correlation between the CMB-ILC reconstruction error $\delta_{\ell m}$ and the instrumental noise $n_{\ell m}^i$) and we express the correlations between the CMB-ILC reconstruction error $\delta_{\ell m}$ and the CMB $s_{\ell m}$ (loss of CMB power due to empirical correlations) by the following formula computed in \cite{2009A&A...493..835D} 
\begin{displaymath}
E\left[ s_{\ell m} \frac{\delta_{\ell m}}{b_\ell^W}\right] = 4\times {(1-\nobs)\over N_p}C_\ell^{\rm CMB} \approx - 0.02 C_\ell^{\rm CMB}, 
\end{displaymath}
where the number of observation frequencies is $\nobs = 6$ and the number of pixels is $N_p = 1024$. 
One finds that
\begin{displaymath}
E_\ell  =  N_\ell^i + \left({\alpha_\ell b_\ell^i\over b_\ell^W}\right)^2 N_\ell^{\rm ILC} + \left[\left(b_\ell^i\left(1-\alpha_\ell\right)\right)^2 + 2b_\ell^i\left(1-\alpha_\ell\right) \left(\alpha_\ell b_\ell^i\right)\times 0.02\right]C_\ell^{\rm CMB},
\end{displaymath}
where we note $N_\ell^i = E\left[\left(n_{\ell m}^i\right)^2\right]$, $N_\ell^{\rm ILC} = E\left[\left(\delta_{\ell m}\right)^2\right]$, and $C_\ell^{\rm CMB}=E\left[s_{\ell m}^2\right]$.
Introducing the following notation for the effect of both Wiener filtering and rebeaming 
\begin{displaymath}
r_\ell =  \left({\alpha_\ell b_\ell^i\over b_\ell^W}\right) = \left({w_\ell b_\ell^i\over (b_\ell^W)^2}\right),
\end{displaymath}
we thus find
\begin{displaymath}  
E_\ell  =  N_\ell^i + r_\ell^2 N_\ell^{\rm ILC} + \left[\left(b_\ell^i\left(1-\alpha_\ell\right)\right)^2 + 2b_\ell^i\left(1-\alpha_\ell\right)b_\ell^W r_\ell \times 0.02 \right]C_\ell^{\rm CMB}.
\end{displaymath}         

\bibliography{biblio}

\begin{figure*}
\begin{center}
\begin{tabular}{c}
\includegraphics[width=125mm,angle=0]{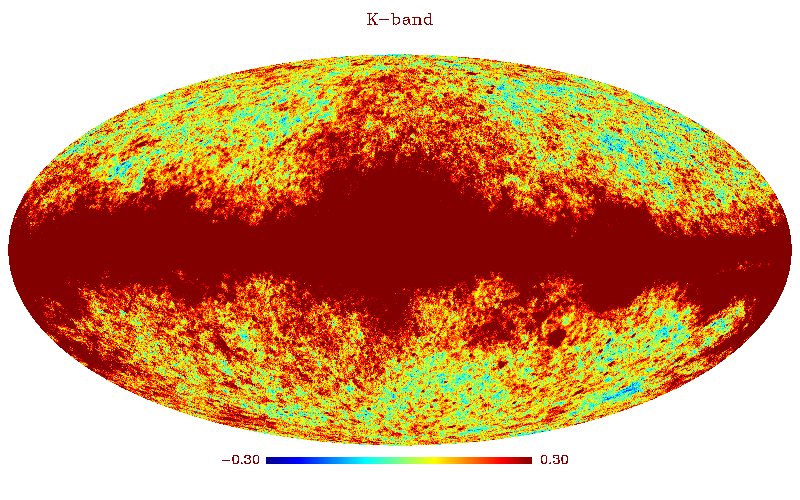}\\
\includegraphics[width=125mm,angle=0]{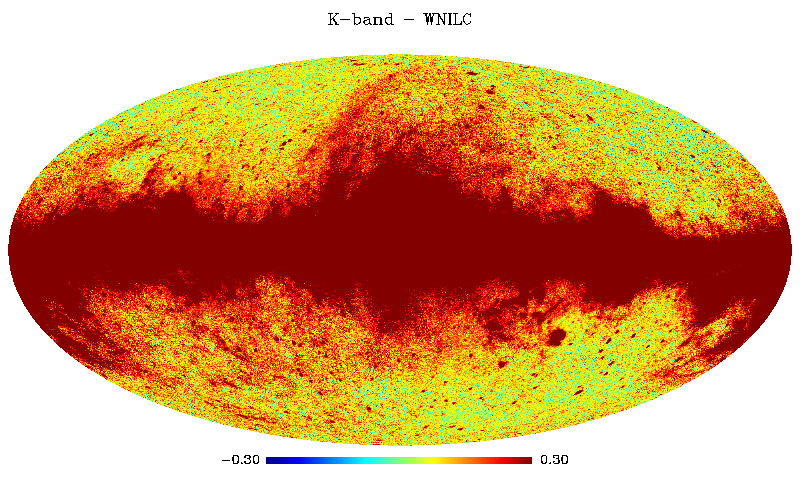}\\
\includegraphics[width=125mm,angle=0]{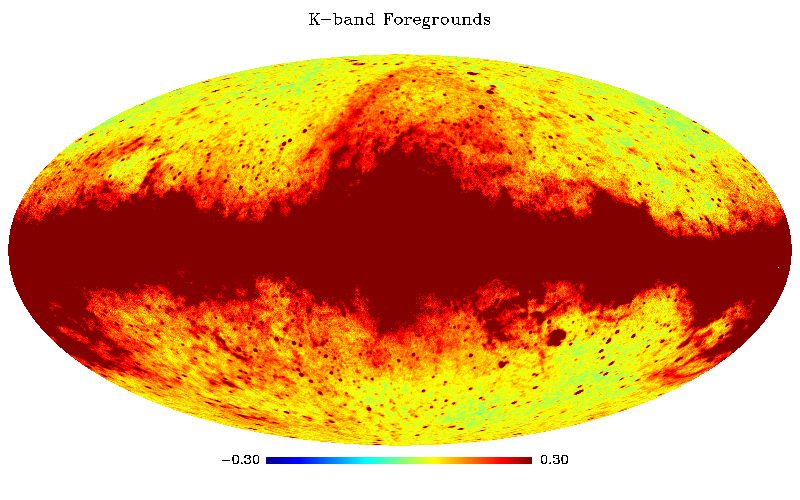}
\end{tabular}
\caption{K-band maps at K-band resolution}
\label{fig:fullskymaps1}
\end{center}
\end{figure*}

\begin{figure*}
\begin{center}
\begin{tabular}{c}
\includegraphics[width=125mm,angle=0]{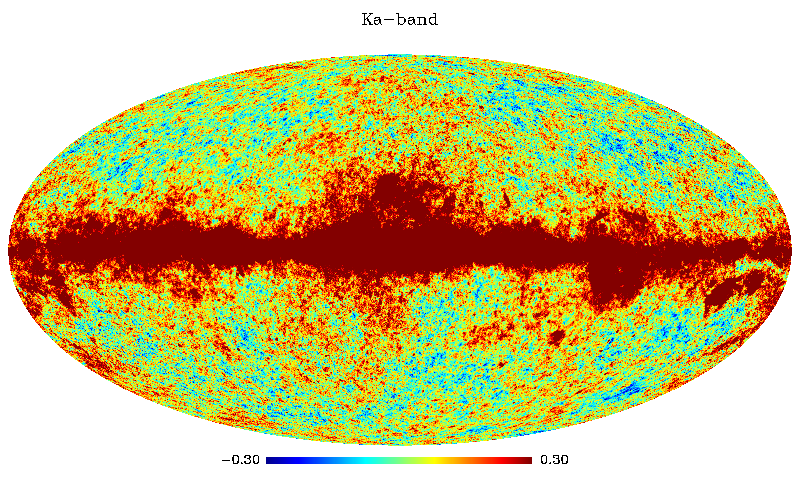}\\
\includegraphics[width=125mm,angle=0]{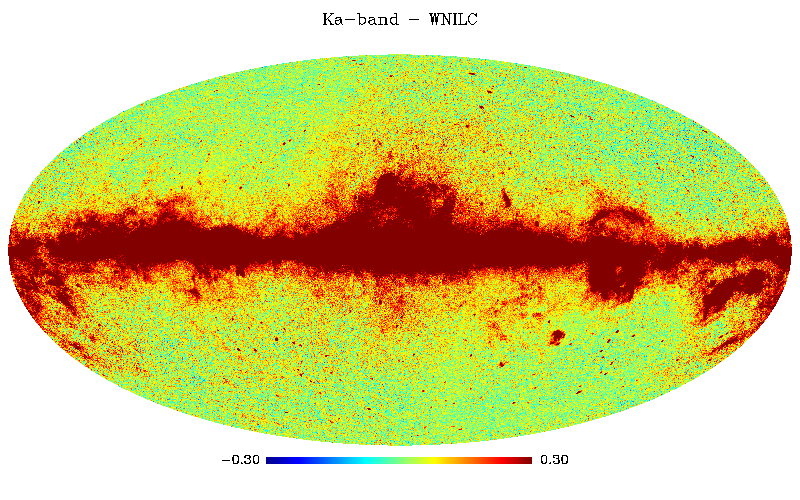}\\
\includegraphics[width=125mm,angle=0]{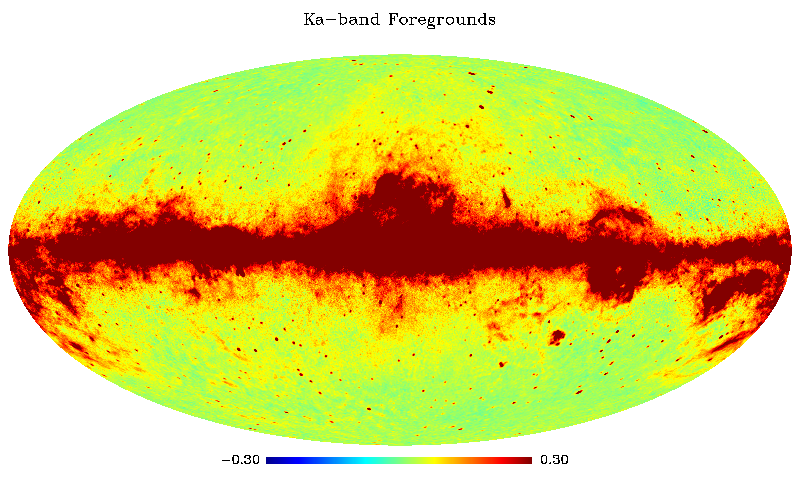}
\end{tabular}
\caption{Ka-band maps at Ka-band resolution}
\label{fig:fullskymaps2}
\end{center}
\end{figure*}

\begin{figure*}
\begin{center}
\begin{tabular}{c}
\includegraphics[width=125mm,angle=0]{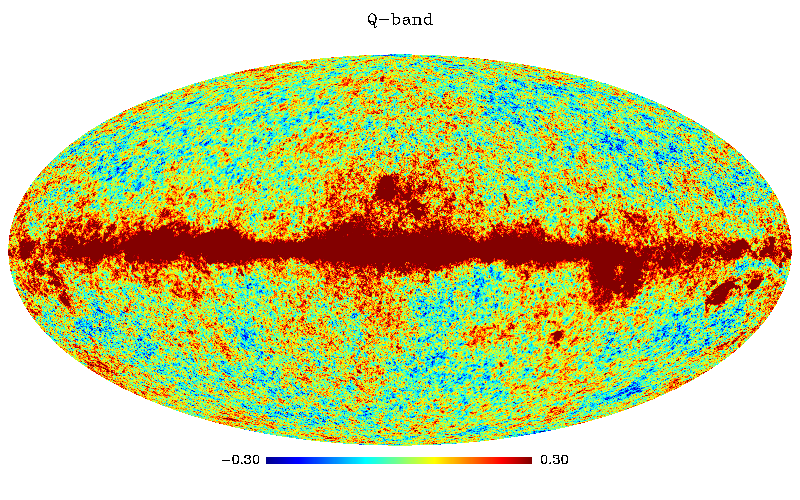}\\
\includegraphics[width=125mm,angle=0]{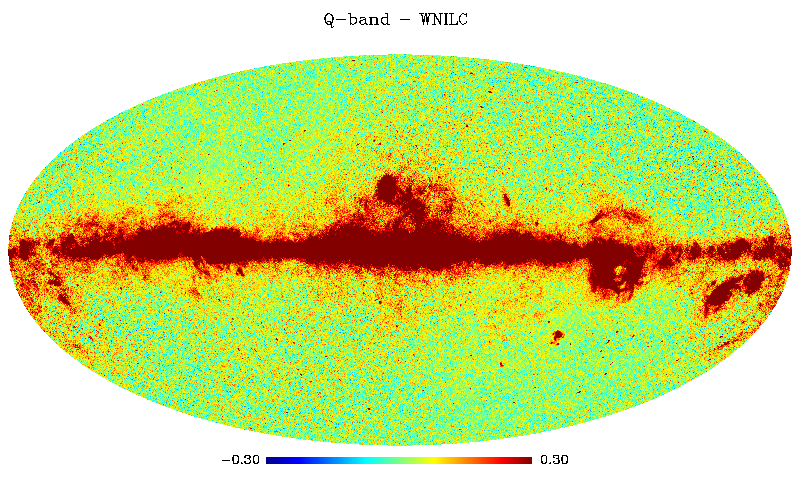}\\
\includegraphics[width=125mm,angle=0]{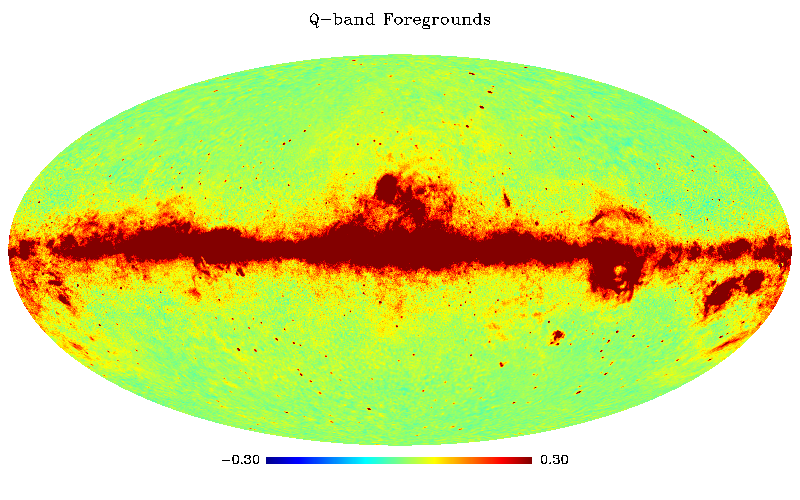}
\end{tabular}
\caption{Q-band maps at Q-band resolution}
\label{fig:fullskymaps3}
\end{center}
\end{figure*}

\begin{figure*}
\begin{center}
\begin{tabular}{c}
\includegraphics[width=125mm,angle=0]{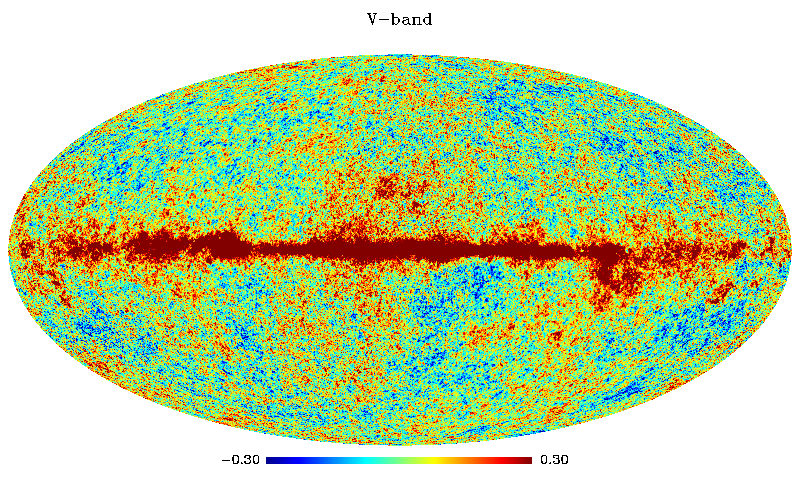}\\
\includegraphics[width=125mm,angle=0]{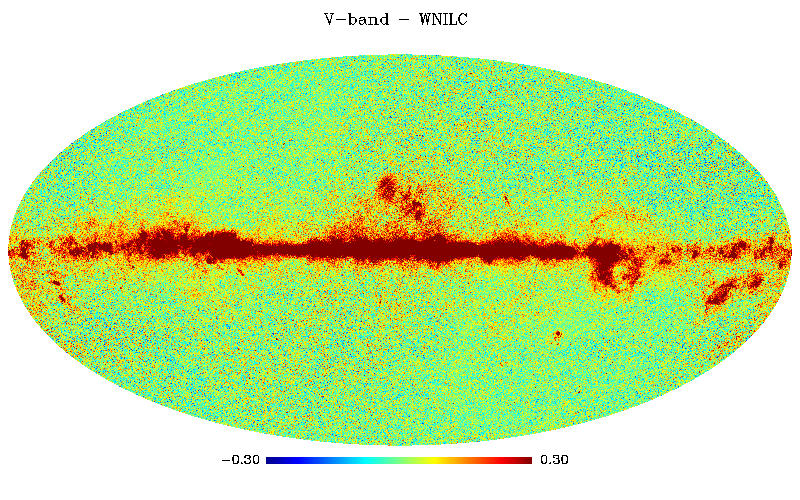}\\
\includegraphics[width=125mm,angle=0]{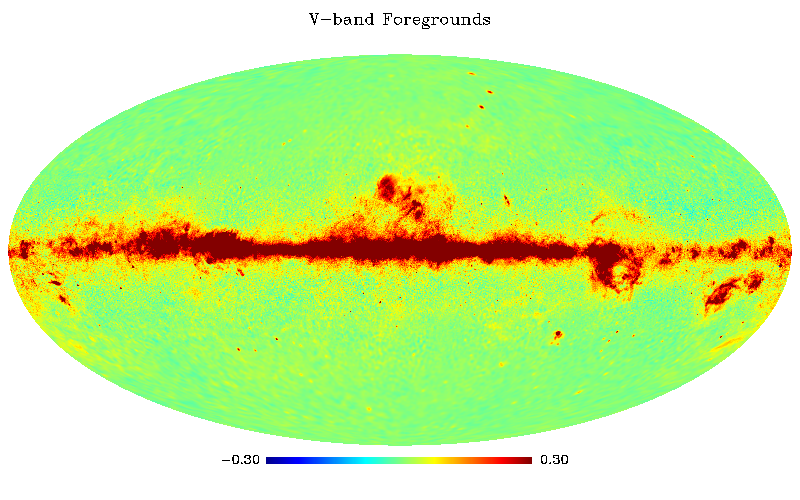}
\end{tabular}
\caption{V-band maps at V-band resolution}
\label{fig:fullskymaps4}
\end{center}
\end{figure*}

\begin{figure*}
\begin{center}
\begin{tabular}{c}
\includegraphics[width=125mm,angle=0]{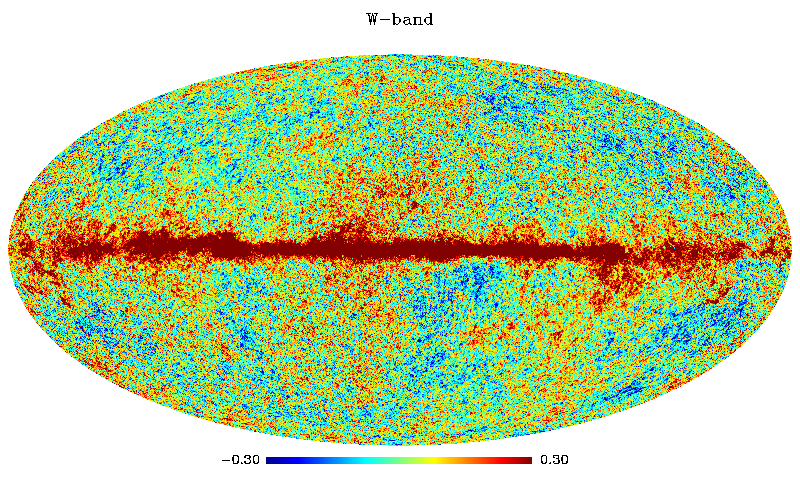}\\
\includegraphics[width=125mm,angle=0]{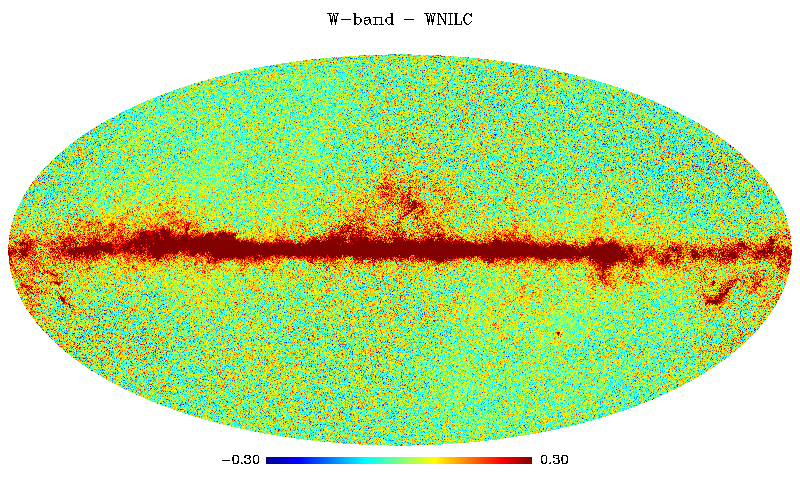}\\
\includegraphics[width=125mm,angle=0]{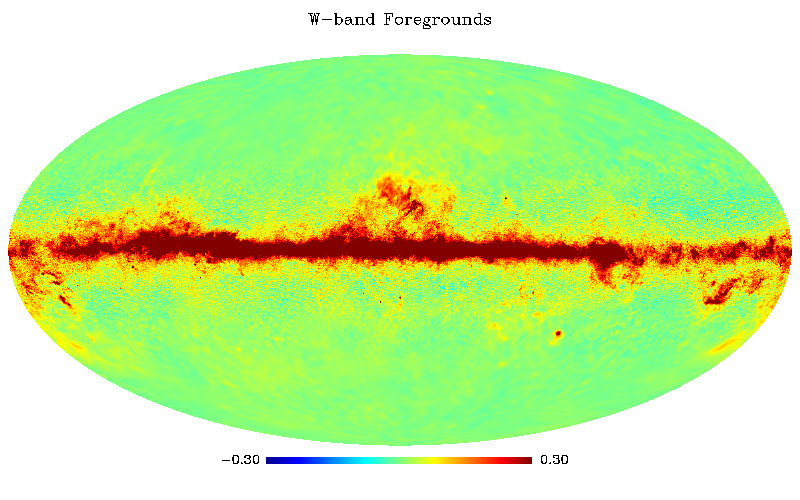}
\end{tabular}
\caption{W-band maps at W-band resolution}
\label{fig:fullskymaps5}
\end{center}
\end{figure*}

\end{document}